\documentclass[
reprint,
superscriptaddress,
nofootinbib,
amsmath,amssymb,
aps,
pra,
floatfix
]{revtex4-2}

\usepackage{lmodern}
\usepackage{graphicx}
\usepackage{bm}
\usepackage{physics}
\usepackage{newtxtext,newtxmath}
\usepackage{color}
\usepackage[usenames,dvipsnames]{xcolor}
\usepackage[breaklinks, pdftex, hyperfootnotes=true, pdfpagelabels, bookmarks, pageanchor]{hyperref}
\usepackage{ulem}

\hypersetup{
	colorlinks=true, linktocpage=true, pdfstartpage=1, pdfstartview=FitH, pdfborder={0 0 0},
	breaklinks=true, pdfpagemode=UseNone, pageanchor=true, pdfpagemode=UseOutlines,
	plainpages=false, bookmarksnumbered, bookmarksopen=true, bookmarksopenlevel=1,
	hypertexnames=true, pdfhighlight=/O,
	urlcolor=blue, linkcolor=blue, citecolor=blue,
}

\graphicspath{{./fig/}}

\begin{document}

\title{Nonstoquastic catalyst for bifurcation-based quantum annealing of ferromagnetic $p$-spin model}

\author{Yuki Susa}
\email{y-susa@nec.com}
\affiliation{Secure System Platform Research Laboratories, NEC Corporation, Kawasaki, Kanagawa 211-8666, Japan}
\affiliation{NEC-AIST Quantum Technology Cooperative Research Laboratory, National Institute of Advanced Industrial Science and Technology (AIST), Tsukuba, Ibaraki 305-8568, Japan}

\author{Takashi Imoto}
\affiliation{Research Center for Emerging Computing Technologies,
National Institute of Advanced Industrial Science and Technology (AIST),
1-1-1 Umezono, Tsukuba, Ibaraki 305-8568, Japan.}

\author{Yuichiro Matsuzaki}
\affiliation{Research Center for Emerging Computing Technologies,
National Institute of Advanced Industrial Science and Technology (AIST),
1-1-1 Umezono, Tsukuba, Ibaraki 305-8568, Japan.}
\affiliation{NEC-AIST Quantum Technology Cooperative Research Laboratory, National Institute of Advanced Industrial Science and Technology (AIST), Tsukuba, Ibaraki 305-8568, Japan}

\date{\today}

\begin{abstract} 
Introducing a nonstoquastic catalyst is a promising avenue to improve quantum annealing with the transverse field.
In the present paper, we propose a nonstoquastic catalyst for bifurcation-based quantum annealing described by the spin-1 operators to improve the efficiency of a ground-state search.
To investigate the effect of the nonstoquastic catalyst, we study the ferromagnetic $p$-spin model, which has difficulty with finding the ground state due to the first-order phase transition for quantum annealing.
A semiclassical analysis shows that the problematic first-order phase transition can be eliminated by introducing the proposed nonstoquastic catalyst with the appropriate amplitude.
We also numerically calculate the minimum energy gap for a finite-size system by diagonalizing the Hamiltonian. We find that while the energy gap decreases exponentially with increasing system size for the original Hamiltonian, it decreases polynomially against the system size for the Hamiltonian with the nonstoquastic catalyst.
This result implies that the proposed nonstoquastic catalyst has the potential to improve the performance of bifurcation-based quantum annealing.
\end{abstract}

\maketitle
\section{Introduction}
Quantum annealing (QA) is a quantum metaheuristic for solving combinatorial optimization problems \cite{kadowaki1998quantum, brooke1999quantum, santoro2002theory, santoro2006optimization, das2008colloquium, morita2008mathematical, hauke2020perspectives}, and is related to adiabatic quantum computation \cite{farhi2001quantum, albash2018adiabatic}.
The target of QA is the ground state of the classical Ising model to which a combinatorial optimization problem is mapped \cite{lucas2014ising}.
Standard QA is formulated as a spin-1/2 Ising model with a time-dependent transverse field.
The protocol of QA starts with the spins initialized to the superposition of the two orthogonal states $\ket{\pm1}$ by the transverse field.
By adiabatically decreasing the amplitude of transverse field, we can obtain the desired ground state corresponding to the optimal solution.

The performance of QA can be evaluated with a minimum energy gap between an instantaneous ground state and the first excited state.
This can be understood with the quantum adiabatic theorem, which indicates that the annealing time necessary to obtain the desired ground state is inversely proportional to the square of the minimum energy gap \cite{jansen2007bounds, amin2009consistency, lidar2009adiabatic}.
It is empirically known that the minimum energy gap closes exponentially with increasing system size when a quantum system encounters a first-order phase transition during annealing. This is a serious problem for QA because the annealing time increases exponentially as the system size increases.
The ferromagnetic $p$-spin model is a well-known example in which the first-order phase transition appears during QA \cite{jorg2010energy}.
In contrast, when the phase transition is second order, the minimum energy gap decreases polynomially as the system size increases. Then, in this case, we can find the ground state in polynomial time.

It is worth mentioning that standard QA is implemented with a stoquastic Hamiltonian, in which all nondiagonal elements of the matrix representation are real and nonpositive \cite{bravyi2008complexity}.
Since a system under a stoquastic Hamiltonian can be emulated classically without the sign problem, the standard QA is considered to have comparable performance to a classical algorithm.
Therefore, whether a nonstoquastic catalyst, which is an additional Hamiltonian violating the stoquastic condition, can improve the performance of QA has been investigated.
The mean-field analysis for QA with the $p$-spin model showed that a certain type of nonstoquastic catalyst is effective in changing the first-order phase transition to the second-order one \cite{seki2012quantum,seoane2012many,nishimori2017exponential,albash2019role}. This is an interesting case in which a nonstoquastic catalyst leads to an exponential acceleration of QA.
The numerical calculation \cite{hormozi2017nonstoquastic} showed that QA under a nonstoquastic Hamiltonian has an advantage over standard QA. 
On the other hand, Ref. \cite{crosson2020signing} showed that the energy gap in a Hamiltonian with a nonstoquastic catalyst is generally smaller than a stoquasticized Hamiltonian obtained by de-signing the nonstoquastic catalyst.

Recently, bifurcation-based QA (BQA) using Kerr-nonlinear parametric oscillators (KPOs) was studied~\cite{milburn1991quantum, wielinga1993quantum, cochrane1999macroscopically, goto2016bifurcation,puri2017engineering,nigg2017robust,puri2017quantum,zhao2018two,goto2019quantum,goto2020quantum,kanao2021high,yamaji2022development}.
A KPO can generate a cat state (superposition of two coherent states) from a vacuum state by an adiabatic process. 
The idea of BQA originates from this cat-state generation process, which is referred to as the quantum bifurcation mechanism.
While the Hamiltonian of BQA using the KPOs is described with bosonic operators, a spin formulation of BQA, which is described by the spin-1 operators, was also proposed in \cite{takahashi2022bifurcation}.
This spin formulation is designed to resemble the bifurcation mechanism of the KPO.
In this formulation, each spin state is initially prepared in $\ket{0}$ and eventually becomes either $\ket{+1}$ or $\ket{-1}$.
According to the effective spin model of the KPO studied in Ref. \cite{miyazaki2022effective}, BQA described with spin-1 operators can be regarded as the approximation model of BQA using KPO.
Also, by adopting a spin-locking technique~\cite{loretz2013radio,doherty2013nitrogen}, we can implement not only the conventional QA \cite{matsuzaki2020quantum,imoto2022obtaining} but also BQA described with the spin-1 operators with nitrogen-vacancy (NV) centers in diamonds \cite{matsuzaki2022generation}.
The NV center in a diamond is a promising device for realizing quantum information processing because it has a long coherence time, such as a few milliseconds, even at room temperature~\cite{balasubramanian2009ultralong,herbschleb2019ultra}.
Therefore, the study of BQA will lead to the development of a novel experimental platform for QA.
In addition, some studies suggest that BQA may have an advantage over conventional QA with the transverse field \cite{puri2017quantum, takahashi2022bifurcation}.
However, the Hamiltonian of the spin formulation of BQA proposed in Ref. \cite{takahashi2022bifurcation} is stoquastic.
The natural question of whether a nonstoquastic catalyst will benefit BQA arises.

In this paper, we study the effect of a nonstoquastic catalyst in BQA described by the spin-1 operators.
Here, we consider the $p$-spin model and propose a nonstoquastic catalyst to change the order of the phase transition.
In order to analyze the phase transitions in stoquastic and nonstoquastic cases, we calculate the energy potential with the semiclassical approximation as in various QA studies \cite{farhi2002quantum1,farhi2002quantum2,schaller2010role,boixo2016computational,muthukrishnan2016tunneling,susa2017relation}.
A previous study \cite{susa2017relation} showed that, for standard QA of the $p$-spin model with the nonstoquastic catalyst, the semiclassical analysis significantly predicts the location and order of the phase transitions, and agrees with the full quantum statistical-mechanical calculations \cite{seki2012quantum}.
Thus, the semiclassical analysis will also be practical for the current problem. We show that the proposed nonstoquastic catalyst is effective for changing the first-order phase transition, which appears in the stoquastic case, to the second-order one.
To support the argument of the semiclassical analysis, we study the exact ground state and the minimum energy gap in a finite-size system by diagonalizing the Hamiltonian.
We confirm that the instantaneous ground state obtained in the semiclassical analysis approximates the exact instantaneous ground state well.
We also show that, as the system size increases, the minimum energy gap decreases exponentially in the stoquastic case and decreases polynomially in the nonstoquastic case.

This paper is organized as follows.
In Sec. \ref{sec:2}, we introduce the spin formulation of BQA.
In Sec. \ref{sec:3}, we formulate the Hamiltonian of the $p$-spin model and propose a nonstoquastic catalyst for BQA.
In Sec. \ref{sec:4}, we calculate the semiclassical potential and the order parameters to investigate the order of the phase transitions.
In Sec. \ref{sec:5}, we discuss the effect of the proposed nonstoquastic catalyst in
a finite-size system.
We summarize this paper in Sec. \ref{sec:6}.
In Appendix \ref{app:A}, we discuss an implementation of BQA with the NV centers in diamonds.
We provide additional analyses and details of certain calculations in Appendixes \ref{app:B} and \ref{app:C}, respectively.

\section{Review of the spin formulation of bifurcation-based quantum annealing}
\label{sec:2}
We recapitulate the spin formulation of BQA proposed in Ref. \cite{takahashi2022bifurcation}.
First, we define the $x$, $y$, and $z$ components of the spin-1 operators as
\begin{subequations}
\begin{align}
\hat{S}^x& =\frac{1}{\sqrt{2}}
\begin{pmatrix}
0&1&0\\
1&0&1\\
0&1&0
\end{pmatrix}, \\
\hat{S}^y&=\frac{i}{\sqrt{2}}
\begin{pmatrix}
0&-1&0\\
1&0&-1\\
0&1&0
\end{pmatrix}, \\
\hat{S}^z&=
\begin{pmatrix}
1&0&0\\
0&0&0\\
0&0&-1
\end{pmatrix},
\end{align}
\end{subequations}
respectively.
The eigenstates of $\hat{S}^z$ denote $\ket{0}$ and $\ket{\pm1}$ as $\hat{S}^z\ket{m} = m \ket{m}$ for $m=0,\pm 1$.
The Hamiltonian of BQA for an $N$-spin system is given by
\begin{align}
\label{eq:1}
\hat{H}(s) = \sum_{i=1}^N \qty[-A(s)\hat{S}_i^x-B(s)(\hat{S}_i^z)^2]+\hat{H}_p(\qty{\hat{S}_i^z}),
\end{align}
where $s\in [0,1]$ is a dimensionless time parameter and $i$ indicates the index of a spin site.
$A(s)$ is a positive function that takes a finite value at the middle of annealing, and $B(s)$ is a function increasing from a large negative value to a large positive value as time $s$ evolves.
We assume that $A(s)$ and $B(s)$ are a Gaussian function and a linear function, respectively, as
\begin{subequations}
\label{eq:3}
\begin{align}
A(s)&:=A_0\exp(-\frac{(2s-1)^2}{2\sigma^2}), \\
B(s)&:=B_0(2s-1).
\end{align}
\end{subequations}
The summation term in Eq. (\ref{eq:1}) is a driver Hamiltonian inducing the bifurcation mechanism. $\hat{H}_p$ represents a problem Hamiltonian that includes interactions between spins and local fields. 
BQA starts with the state $\otimes_i \ket{0}_i$, which is the ground state of the Hamiltonian (\ref{eq:1}) at $s=0$. By adiabatically evolving the system under the Hamiltonian (\ref{eq:1}) from $s=0$ to $s=1$, each spin state changes to $\ket{\pm1}_i$, corresponding to the ground state of $\hat{H}_p$ that is our target.

\begin{figure}[tbp]
 \centering
	\includegraphics[width = 8.33 cm]{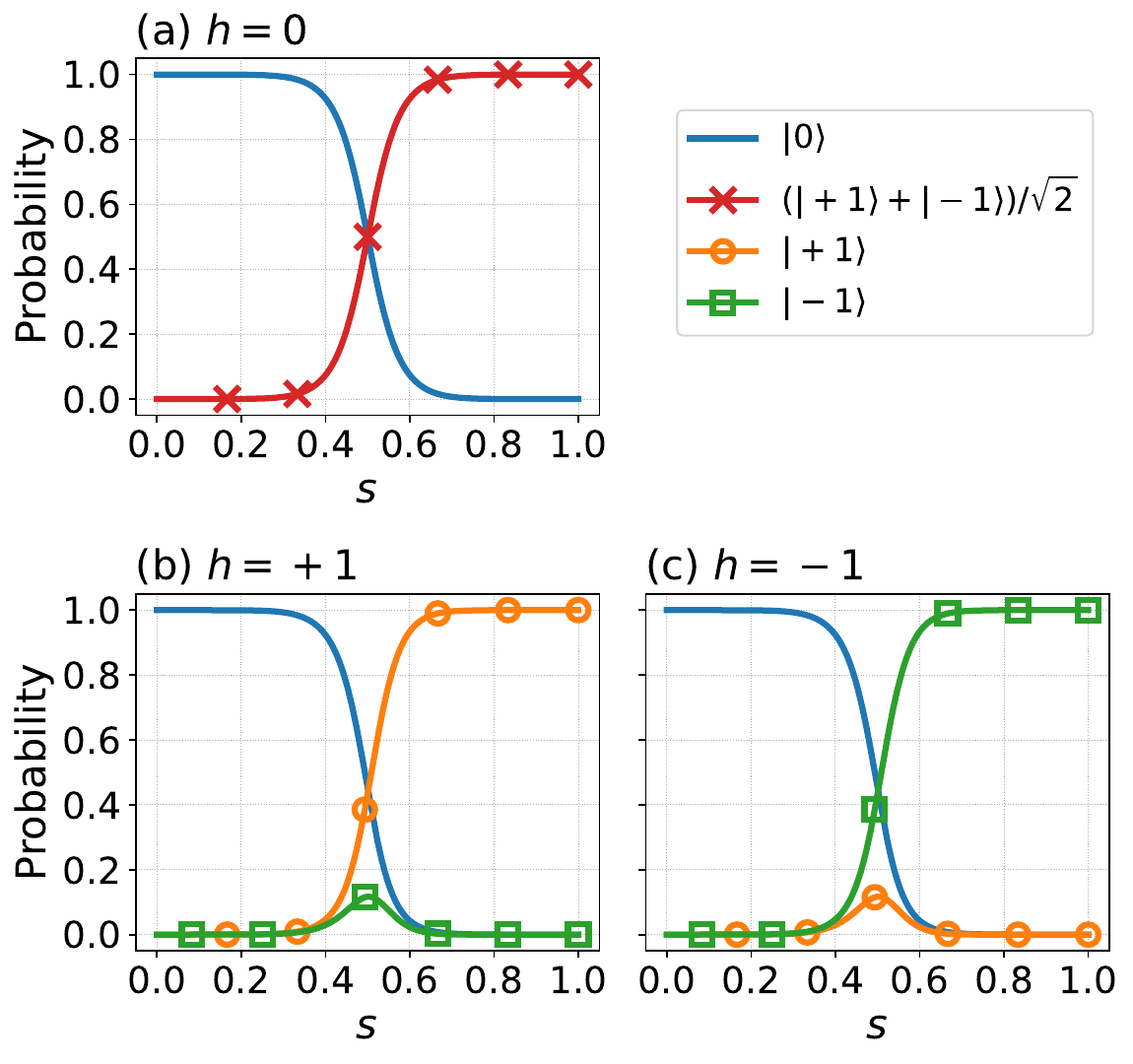}
\caption{Plots of probabilities of each state in the instantaneous ground state of Hamiltonian (\ref{eq:4}) as a function of $s$ for three cases: (a) $h=0$, (b) $h=+1$, and (c) $h=-1$. $A(s)$ and $B(s)$ are from Eq. (\ref{eq:3}) with $A_0 = 3$, $\sigma^2=0.1$, and $B_0 = 40$.}
\label{fig:1}
\end{figure}

To briefly review the bifurcation mechanism, we consider a single-spin system. 
The Hamiltonian is
\begin{align}
\label{eq:4}
\hat{H}^{(1)}(s)=-A(s)\hat{S}^x-B(s)(\hat{S}^z)^2-h\hat{S}^z.
\end{align}
We evaluate the instantaneous ground state numerically for the three $h$ cases.
Figure \hyperref[fig:1]{1(a)} for $h = 0$ shows that the final ground state is the superposition of $\ket{\pm 1}$.
We can see the bifurcation mechanism around $s=0.5$, where the probability of the superposition state increases and the state $\ket{0}$ vanishes.
This case corresponds to the cat-state generation of the KPO.
When $h$ is positive (negative), the spin state becomes $\ket{+1}$ ($\ket{-1}$), as shown in Figs. \hyperref[fig:1]{1(b)} and \hyperref[fig:1]{1(c)}.

\section{Ferromagnetic $p$-spin model and nonstoquastic catalyst}
\label{sec:3}

Hereafter, we consider the ferromagnetic $p$-spin model in the spin formulation of BQA.
The Hamiltonian is as follows:
\begin{align}
\label{eq:5}
\hat{H}(s) = \sum_{i=1}^N \qty[-A(s)\hat{S}_i^x-B(s)(\hat{S}_i^z)^2] -N\qty(\frac{1}{N}\sum_{i=1}^N\hat{S}_i^z)^p.
\end{align}
For odd $p$, the ground state of the $p$-spin model is $\otimes_{i=1}^N \ket{+1}_i$, and for even $p$, the ground state is doubly degenerate, $\otimes_{i=1}^N \ket{+1}_i$ and $\otimes_{i=1}^N \ket{-1}_i$. 
Note that this $p$-spin model reduces to the Grover problem for the limit of $p \rightarrow \infty$. 

This Hamiltonian (\ref{eq:5}) is stoquastic, and the first-order phase transition appears during the time evolution (we discuss this in the next section).
We then propose the following interaction as a nonstoquastic catalyst:
\begin{align}
\label{eq:6}
&\hat{H}_{\mathrm{c}}=N\qty(\frac{1}{N}\sum_{i=1}^N (\hat{S}_i^x)^2-(\hat{S}_i^y)^2 )^2, \\
\label{eq:7}
&(\hat{S}_i^x)^2-(\hat{S}_i^y)^2 =
\begin{pmatrix}
0&0&1\\
0&0&0\\
1&0&0
\end{pmatrix}_i.
\end{align}
The operator (\ref{eq:7}) switches the spin state between $\ket{+1}_i$ and $\ket{-1}_i$ in a way similar to the Pauli $X$ operator.
The catalyst (\ref{eq:6}) is inspired by a nonstoquastic $XX$ interaction removing the first-order phase transition in standard QA of the $p$-spin model \cite{seki2012quantum,seoane2012many,nishimori2017exponential}.
We consider the following Hamiltonian:
\begin{align}
\label{eq:8}
\hat{H}(s)=&\sum_{i=1}^N \qty[-A(s)\hat{S}_i^x-B(s)(\hat{S}_i^z)^2] \notag \\
&+CN\qty(\frac{1}{N}\sum_{i=1}^N (\hat{S}_i^x)^2-(\hat{S}_i^y)^2 )^2 -N\qty(\frac{1}{N}\sum_{i=1}^N\hat{S}_i^z)^p,
\end{align}
where $C$ is the amplitude of the proposed nonstoquastic catalyst.
The Hamiltonian (\ref{eq:8}) becomes nonstoquastic for $C>0$, and setting $C<0$ corresponds to the de-signed stoquastization studied in Ref. \cite{crosson2020signing}.
We discuss a possible realization of this Hamiltonian by using NV centers in diamonds in Appendix \ref{app:A}.
It is worth mentioning that the typical energy scale of the interaction between the NV centers is tens of kilohertz when we realize our proposed method with the NV centers in diamonds, as we discuss in Appendix \ref{app:A}.
Therefore, throughout this paper, we assume that the energy is scaled by a unit of
$10$ kHz.

\section{Semiclassical analysis with spin coherent state}
\label{sec:4}

To investigate the phase transitions in BQA under the Hamiltonian (\ref{eq:8}), we use the semiclassical spin coherent state for the spin-1 operators \cite{runeson2020generalized} defined as the product state
\begin{align}
\label{eq:9}
\ket{\psi_{\textrm{SC}}(\theta,\phi,\alpha,\beta)} =&\otimes_{i=1}^{N}\left[\cos \frac{\theta}{2} \ket{0}_i+\sin\frac{\theta}{2} \cos\frac{\phi}{2} e^{i\alpha}\ket{+1}_i \right. \notag \\
& \left.+\sin\frac{\theta}{2} \sin\frac{\phi}{2} e^{i\beta}\ket{-1}_i \right].
\end{align}
All spins are assumed to have the same angular variables $\theta$, $\phi$, $\alpha$, and $\beta$. 
The semiclassical potential per spin is derived from the expectation value of the Hamiltonian in the spin coherent state (\ref{eq:9}) as
\begin{align}
\label{eq:10}
&V_{\textrm{SC}}(s,\theta,\phi,\alpha,\beta) \notag \\
=& \lim_{N \rightarrow \infty} \frac{1}{N}\bra{\psi_{\textrm{SC}}}\hat{H}(s)\ket{\psi_{\textrm{SC}}} \notag \\
=&-\frac{A(s)}{\sqrt{2}}\sin\theta\qty( \cos\frac{\phi}{2}\cos\alpha+\sin\frac{\phi}{2}\cos\beta) \notag \\
&-B(s)\sin^2\frac{\theta}{2} +C\qty(\sin^2\frac{\theta}{2}\sin \phi \cos(\alpha-\beta) )^2 \notag \\
&- \qty(\sin^2\frac{\theta}{2} \cos\phi)^{p}.
\end{align}
The $\theta$, $\phi$, $\alpha$, and $\beta$ to minimize the semiclassical potential are denoted as $\theta_{\min}$, $\phi_{\min}$, $\alpha_{\min}$, and $\beta_{\min}$, respectively.
The ground state in the semiclassical approximation is given by $\ket{\psi_{\textrm{SC,GS}}}=\ket{\psi_{\textrm{SC}}(\theta_{\min}, \phi_{\min}, \alpha_{\min},\beta_{\min}) }$. Thus, the order parameter at the semiclassical limit can be calculated as
\begin{align}
m := \bra{\psi_{\textrm{SC,GS}}}\frac{1}{N}\sum_{i=1}^{N}\hat{S}_i^z\ket{\psi_{\textrm{SC,GS}}} =\sin^2\frac{\theta_{\min}}{2}\cos\phi_{\min}.
\end{align}

\begin{figure}[tbp]
\centering
  \includegraphics[width = 0.48\textwidth]{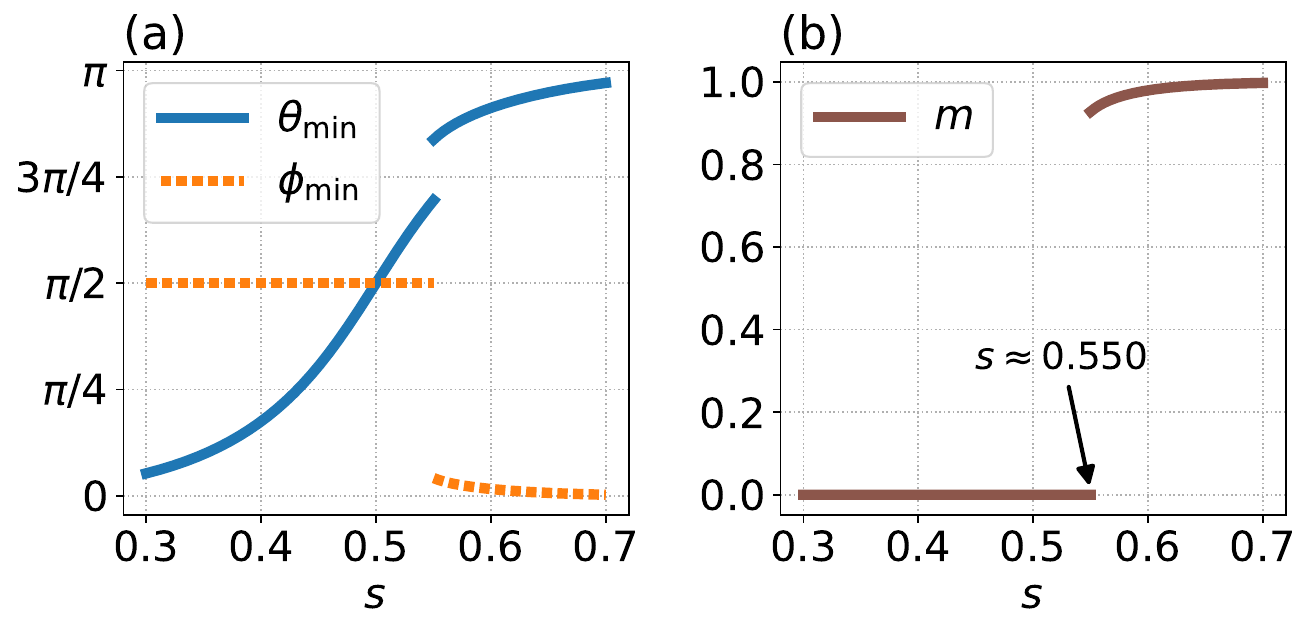}
\caption{Plots of (a) $\theta_{\min}$ and $\phi_{\min}$ and (b) order parameter $m$ against $s$ for $p=5$ in the stoquastic case $C=0$. The first-order phase transition appears at $s\approx 0.550$. $A(s)$ and $B(s)$ are from Eq. (\ref{eq:3}), with $A_0 = 3$, $\sigma^2=0.1$, and $B_0 = 40$.}
\label{fig:2}
\end{figure}

Note that since the spin coherent state (\ref{eq:9}) does not cover superposition of $\otimes_i \ket{+1}_i$ and $\otimes_i \ket{-1}_i$, the present analysis might be insignificant for even $p$ cases with the degenerate ground state.
For convenience, we arbitrarily consider the case where the instantaneous ground state finally becomes $\otimes_i \ket{+1}_i$.
Therefore, we restrict the domains of $\theta$ and $\phi$ to $0\leq \theta \leq \pi$ and $0 \leq \phi \leq \pi/2$.

First, we consider the stoquastic case $C=0$, where $\alpha=\beta=0$ clearly gives the ground state. Figure \ref{fig:2} shows the numerical results for $\theta_{\min}$, $\phi_{\min}$, and the order parameter $m$ for $p=5$.
Figure \hyperref[fig:2]{2(a)} shows $\theta_{\min}$ increases from zero with $\phi_{\min}=\pi/2$ in the first half. 
This means that the amplitude of $\ket{0}_i$ in the instantaneous ground state decreases, and those of $\ket{+1}_i$ and $\ket{-1}_i$ increase.
After the discontinuous change in $\theta_{\min}$ and $\phi_{\min}$, we obtain $\theta_{\min}=\pi$ and $\phi_{\min}=0$, which gives $\ket{\psi_{\textrm{SC,GS}}}=\otimes_i \ket{+1}_i$.
We can see in Fig. \hyperref[fig:2]{2(b)} that the order parameter $m$ changes discontinuously at $s\approx 0.550$, where the first-order phase transition appears.

\begin{figure}[th]
\centering
  \includegraphics[width = 0.48\textwidth]{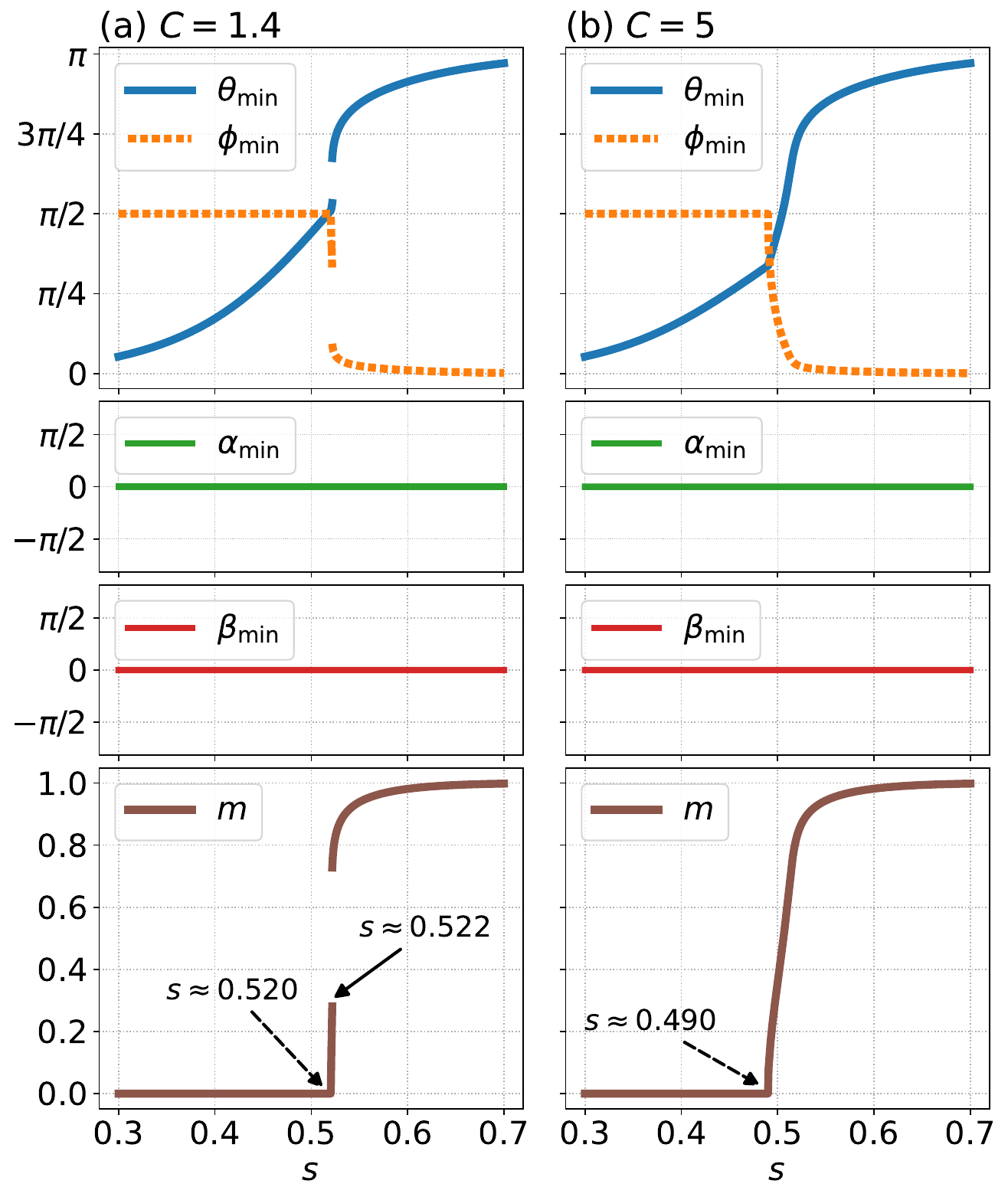}
\caption{Plots of $\theta_{\min}$, $\phi_{\min}$, $\alpha_{\min}$, $\beta_{\min}$, and order parameter $m$ against $s$ for $p=5$ and $C>0$. The first-order phase transition appears at $s\approx0.522$ in (a). The second-order phase transitions are observed at $s\approx 0.520$ in (a) and $s \approx 0.490$ in (b). In both cases, $A(s)$ and $B(s)$ are from Eq. (\ref{eq:3}), with $A_0 = 3$, $\sigma^2=0.1$, and $B_0 = 40$.}
\label{fig:3}
\end{figure}

The results for the nonstoquastic cases are plotted in Fig. \ref{fig:3}.
In Fig. \hyperref[fig:3]{3(a)} for $C=1.4$, $\phi_{\min}$ deviates from $\pi/2$ at $s \approx 0.520$.
Correspondingly, the order parameter is continuously away from zero, which is a signature of a second-order transition. However, the first-order phase transition occurs at $s \approx 0.522$.
By setting a significant amplitude, such as $C = 5$, as shown in Fig. \hyperref[fig:3]{3(b)}, the phase transition becomes completely second order, where the order parameter changes continuously from $0$ to $1$. 
Therefore, in order to resolve the first-order phase transition, the proposed nonstoquastic catalyst (\ref{eq:6}) is effective.
Note that we obtain $\alpha_{\min}=\beta_{\min}=0$ in both Figs. \hyperref[fig:3]{3(a)} and \hyperref[fig:3]{3(b)}.
This is due to the absence of an imaginary component in the Hamiltonian (\ref{eq:8}).
In Appendix \ref{app:B} we give the situation where the driver Hamiltonian is rotated around the $z$ axis and the total Hamiltonian has an imaginary component.

\begin{figure}[bp]
\centering
  \includegraphics[width = 0.48\textwidth]{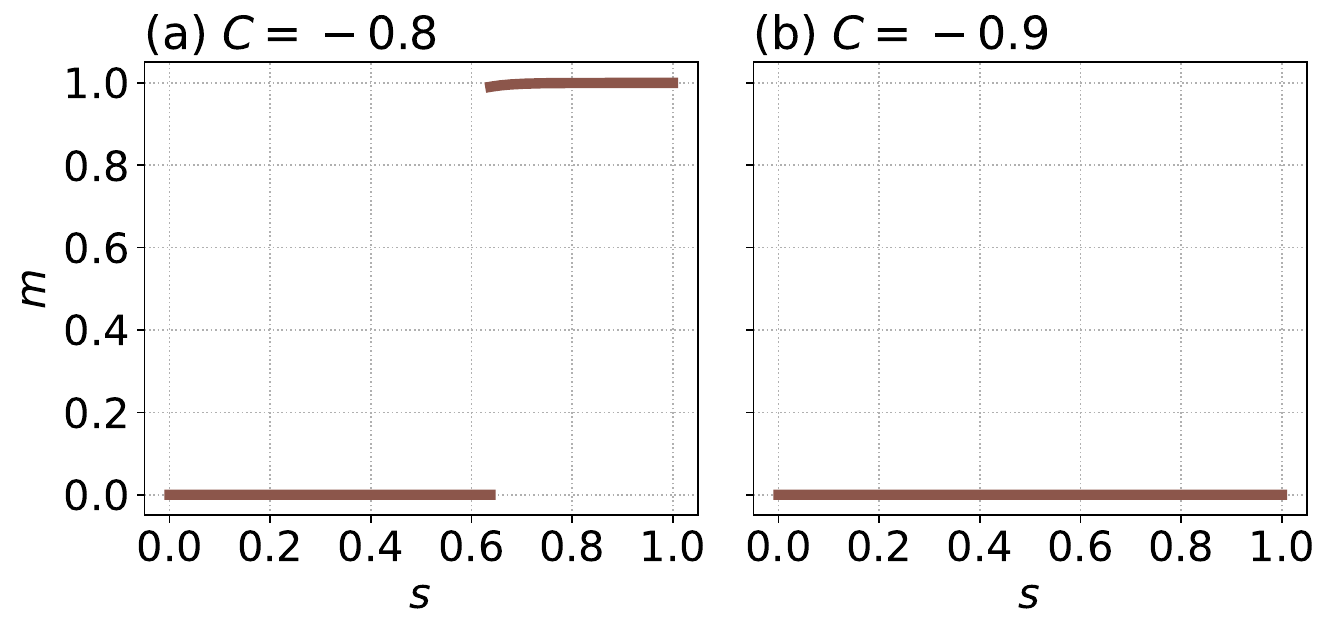}
\caption{Plots of order parameter $m$ against $s$ for $p=5$ and $C<0$. $A(s)$ and $B(s)$ are from Eq. (\ref{eq:3}), with $A_0 = 3$, $\sigma^2=0.1$, and $B_0 = 40$.}
\label{fig:4}
\end{figure}

Next, we discuss de-signed stoquastization \cite{crosson2020signing} of the nonstoquastic Hamiltonian (\ref{eq:8}). In Fig. \ref{fig:4}, we plot the order parameters for two negative-$C$ cases. The first-order phase transition appears when $C=-0.8$ [Fig. \hyperref[fig:4]{4(a)}], and the order parameter $m$ stays zero for $C=-0.9$ [Fig. \hyperref[fig:4]{4(b)}] during time evolution. Thus, negative $C$ is ineffective for the present problem. 

\begin{figure}[tbp]
\centering
  \includegraphics[width= 0.48\textwidth]{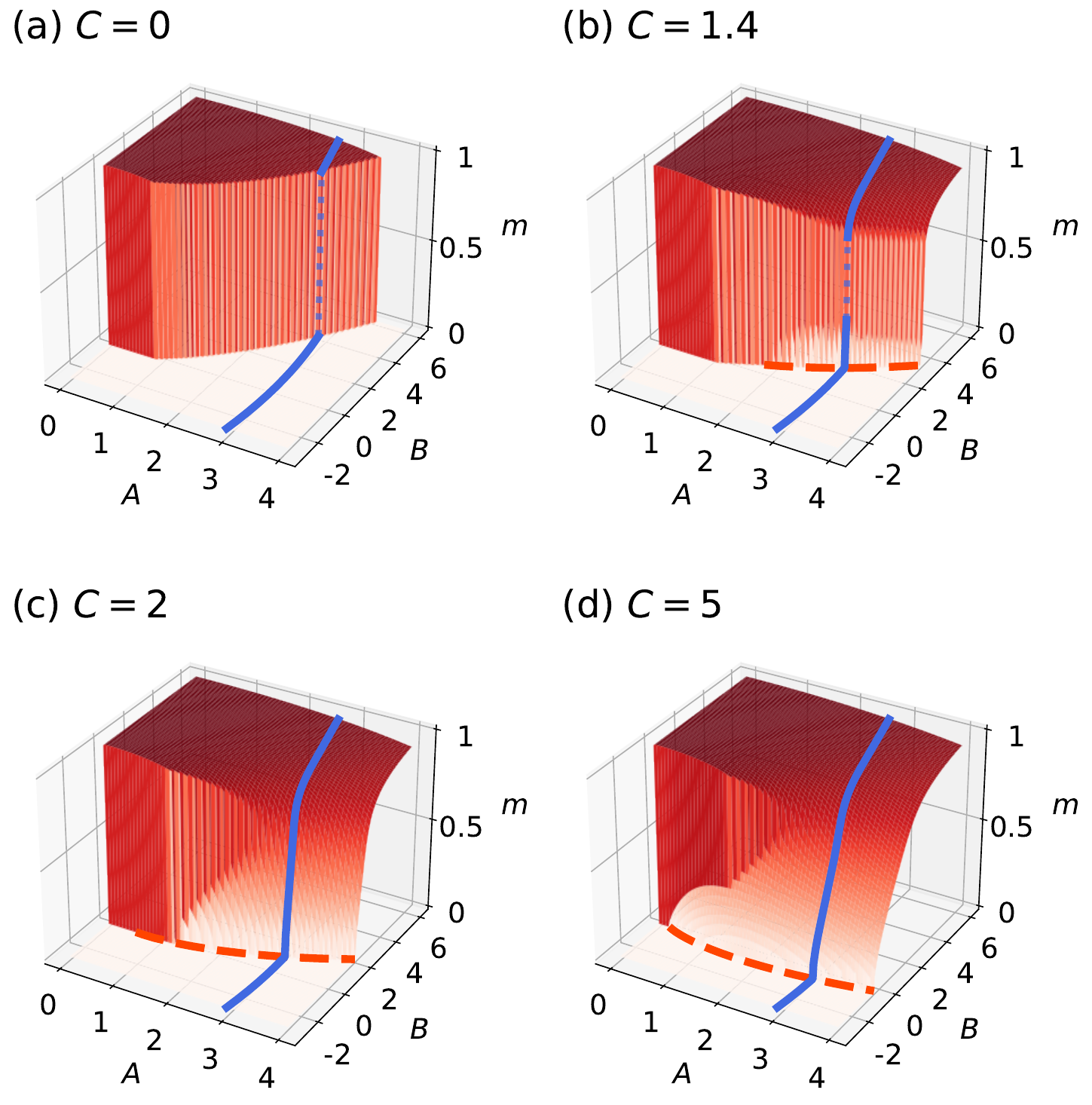}
\caption{Order parameter $m$ as a function of coefficients $A$ and $B$ for $p =5$. The blue curve represents $m$ when one takes the path corresponding to functions $A(s)$ and $B(s)$ in Eq. (\ref{eq:3}) with $A_0 = 3$, $\sigma^2=0.1$, and $B_0 = 40$.
The solid and dotted parts of the blue curves indicate that $m$ continuously and discontinuously changes, respectively. The orange curves in the $A$-$B$ plane indicate the second-order phase transitions given by Eq. (\ref{eq:12}).} 
\label{fig:5}
\end{figure}

We also plot the order parameter as a function of coefficients $A$ and $B$ in the driver Hamiltonian. The results are plotted in Fig. \ref{fig:5}. 
In Fig. \hyperref[fig:5]{5(a)} for the stoquastic case, the system encounters the first-order phase transition by increasing $B$ regardless of the value of $A$.
Next, we show the plots for $C>0$ in Figs. \href{fig:5}{5(b)}-\href{fig:5}{5(d)}.
The orange dashed curves indicate the location where the second-order phase transition appears.
These curves are calculated using the following conditions
\begin{subequations}
\label{eq:12}
\begin{align}
\frac{\partial V_{\mathrm{SC}}}{\partial \theta}\big|_{\phi=\pi/2,\alpha=\beta=0}
&= -A\cos \theta -\frac{B}{2}\sin \theta +C \sin^2 \frac{\theta}{2}\sin \theta \notag \\
&= 0, \\
\frac{\partial^2 V_{\mathrm{SC}}}{\partial \phi^2}\big|_{\phi=\pi/2,\alpha=\beta=0} 
&=\sin \frac{\theta}{2}\qty(\frac{A}{2} \cos \frac{\theta}{2} -2C \sin^3 \frac{\theta}{2}) \notag \\
&=0.
\end{align}
\end{subequations}
In this calculation, we assume that the second-order phase transition occurs at the point where $\phi_{\min}$ continuously deviates from $\pi/2$ and the ground state has $\alpha_{\min}=\beta_{\min} =0$.
Figure \hyperref[fig:5]{5(b)} shows the case of $C=1.4$. The first-order transition appears even if we take the path across the orange dashed curve. The blue curve in Fig. \hyperref[fig:5]{5(b)} is the same as the order parameter plotted in Fig. \hyperref[fig:3]{3(a)}.
In Fig. \hyperref[fig:5]{5(c)}, we can find the path along which the order parameter continuously changes for $C=2$.
However, the first-order phase transition remains for the path with a small $A$.
Although $C$ is made larger, a large $A$ is necessary to avoid the first-order phase transition, as shown in Fig. \hyperref[fig:5]{5(d)}.

\begin{figure}[tbp]
\centering
  \includegraphics[width= 0.48\textwidth]{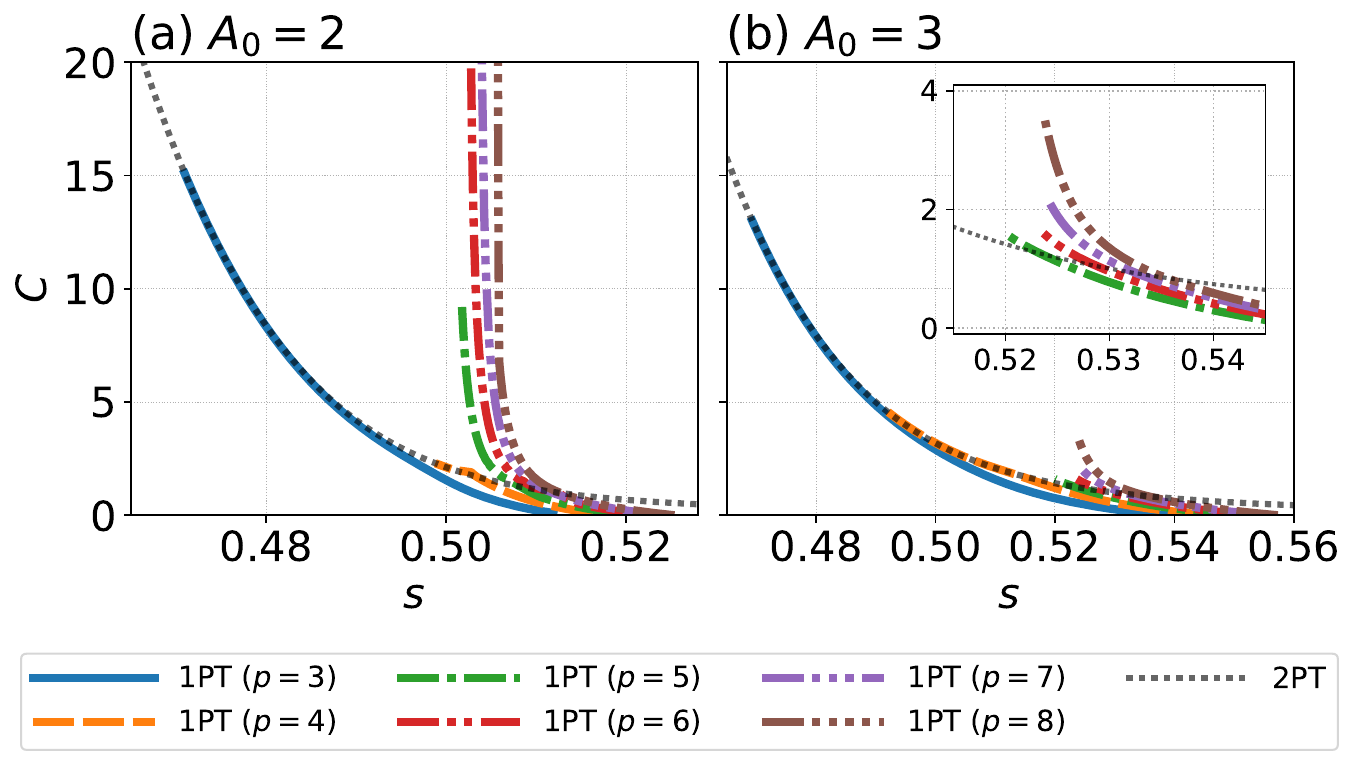}
\caption{Phase diagram in the $s$-$C$ plane for $3 \leq p \leq 8$. $A(s)$ and $B(s)$ are from Eq. (\ref{eq:3}), with $\sigma^2=0.1$ and $B_0 = 40$. Each thick colored curve represents a first-order phase transition (1PT). The black dotted curve indicates the location $s$ of the second-order phase transition (2PT) calculated from Eq. (\ref{eq:12}). } 
\label{fig:6}
\end{figure}

\begin{figure}[bp]
\centering
  \includegraphics[width= 0.45\textwidth]{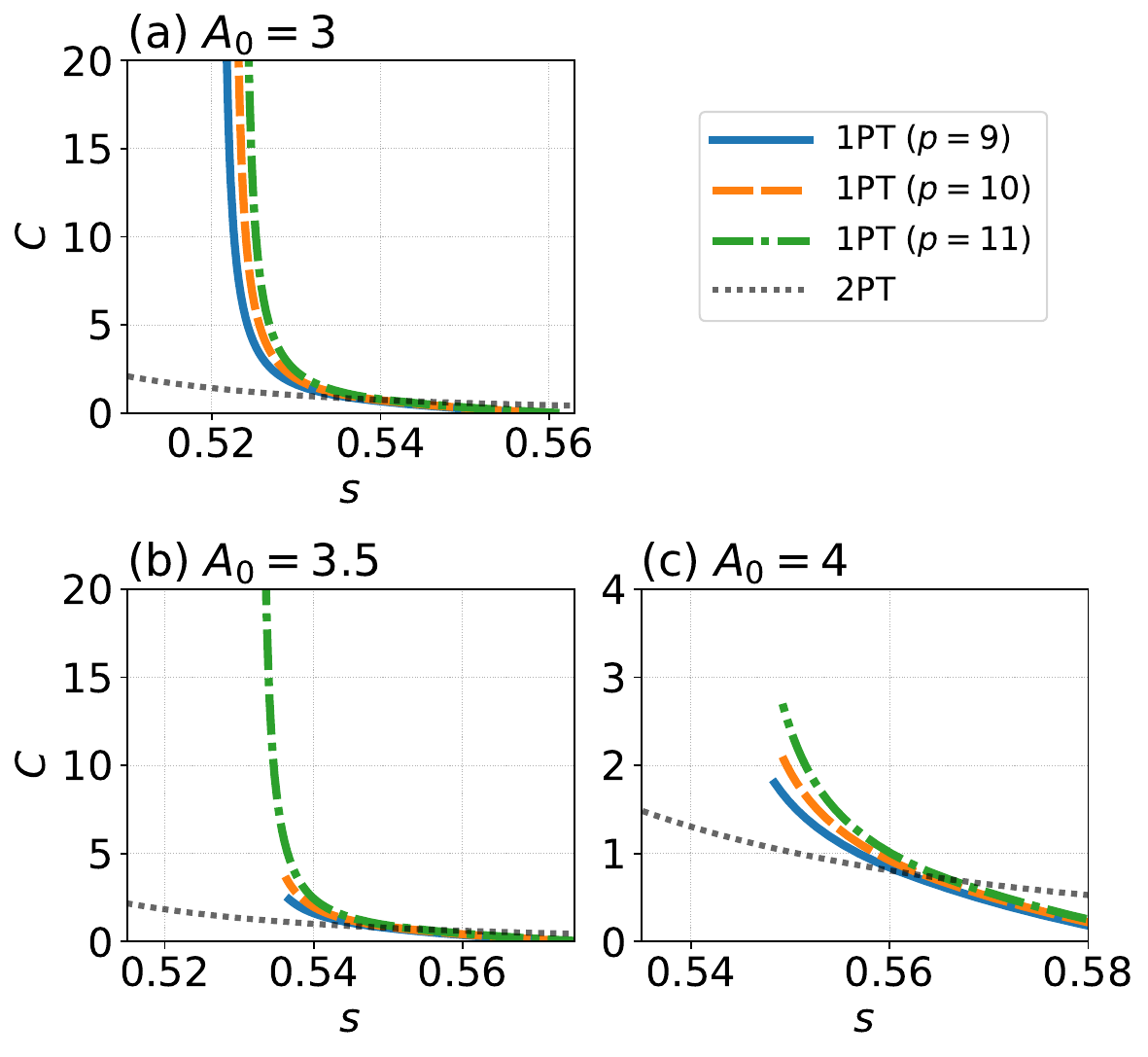}
\caption{Phase diagram in the $s$-$C$ plane for $9 \leq p \leq 11$. $A(s)$ and $B(s)$ are from Eq. (\ref{eq:3}), with $\sigma^2=0.1$ and $B_0 = 40$. Each thick colored curve represents a first-order phase transition (1PT). The black dotted curve indicates the location $s$ of the second-order phase transition (2PT) calculated from Eq. (\ref{eq:12}).}
\label{fig:7}
\end{figure}

Figure \ref{fig:6} shows the phase diagrams for $3 \leq p \leq 8$ obtained from the order parameter, which is given by the semiclassical analysis. We use Eq. (\ref{eq:3}) for $A(s)$ and $B(s)$, and consider two $A_0$ cases.
The colored curves indicating the first-order phase transition extend from points on the axis $C=0$. 
In Fig. \hyperref[fig:6]{6(a)}, the curves for $p=3, 4$, and $5$ are terminated at finite $C$. Therefore, we can find the paths to avoid the first-order phase transitions for $p=3, 4$, and $5$ by tuning the amplitude $C$. However, as long as we use the value of $A_0 = 2$, the curves for the first-order phase transitions for $6 \leq p \leq 8$ are unavoidable.
On the other hand, Fig. \hyperref[fig:6]{6(b)} shows that, if we adopt $A_0=3$, we can circumvent those curves with relatively small $C$. 
However, Fig. \ref{fig:7} shows that we need to set larger $A_0$ again to avoid the first-order phase transition in the higher-$p$ cases. 
It is worth mentioning that for conventional QA with qubits it becomes more difficult to avoid the first-order phase transition as we increase the value of $p$ \cite{seki2012quantum,seoane2012many}, which is similar to our case.
More specifically, we need a careful adjustment of
the amplitude parameter to avoid a first-order phase transition when $p$ is high \cite{seki2012quantum,seoane2012many}.

\section{Analysis for finite-size system}
\label{sec:5}
\begin{figure}[tp]
 \centering
  \includegraphics[width =0.48\textwidth]{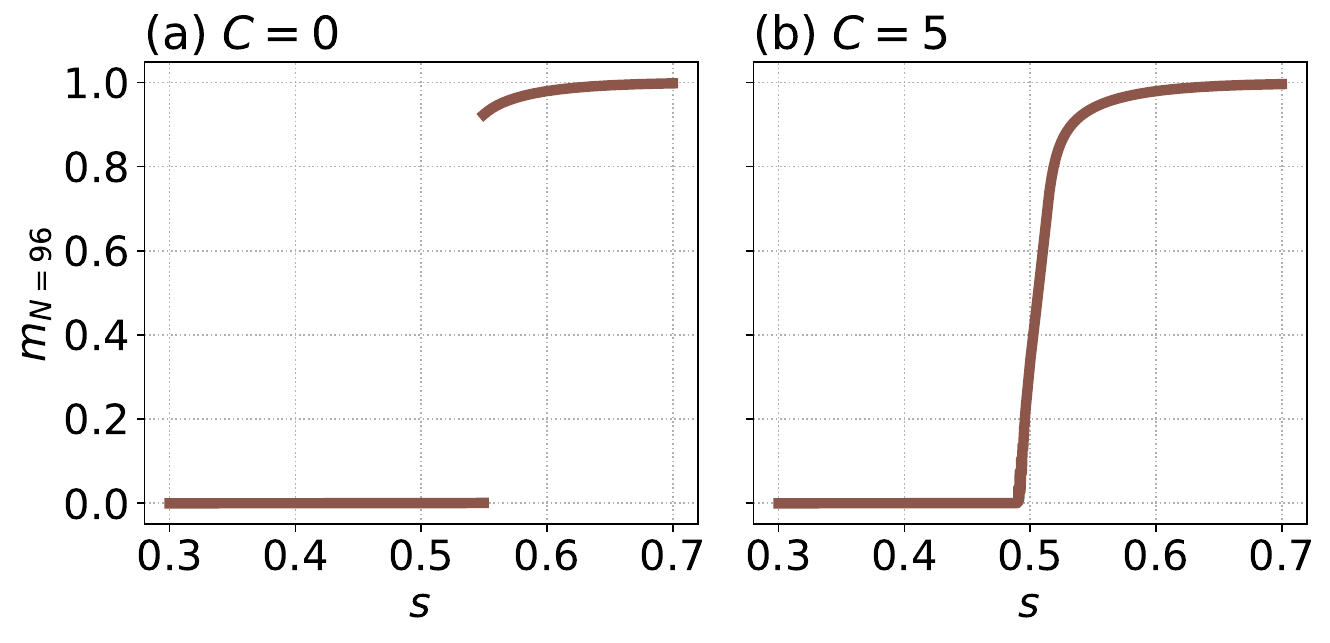}
\caption{Plots of order parameter $m_{N=96}$ against $s$ for $p=5$. $A(s)$ and $B(s)$ are from Eq. (\ref{eq:3}) with $A_0 = 3$, $\sigma^2=0.1$, and $B_0 = 40$.}
\label{fig:8}
\end{figure}

\begin{figure}[bp]
 \centering
  \includegraphics[width =0.48\textwidth]{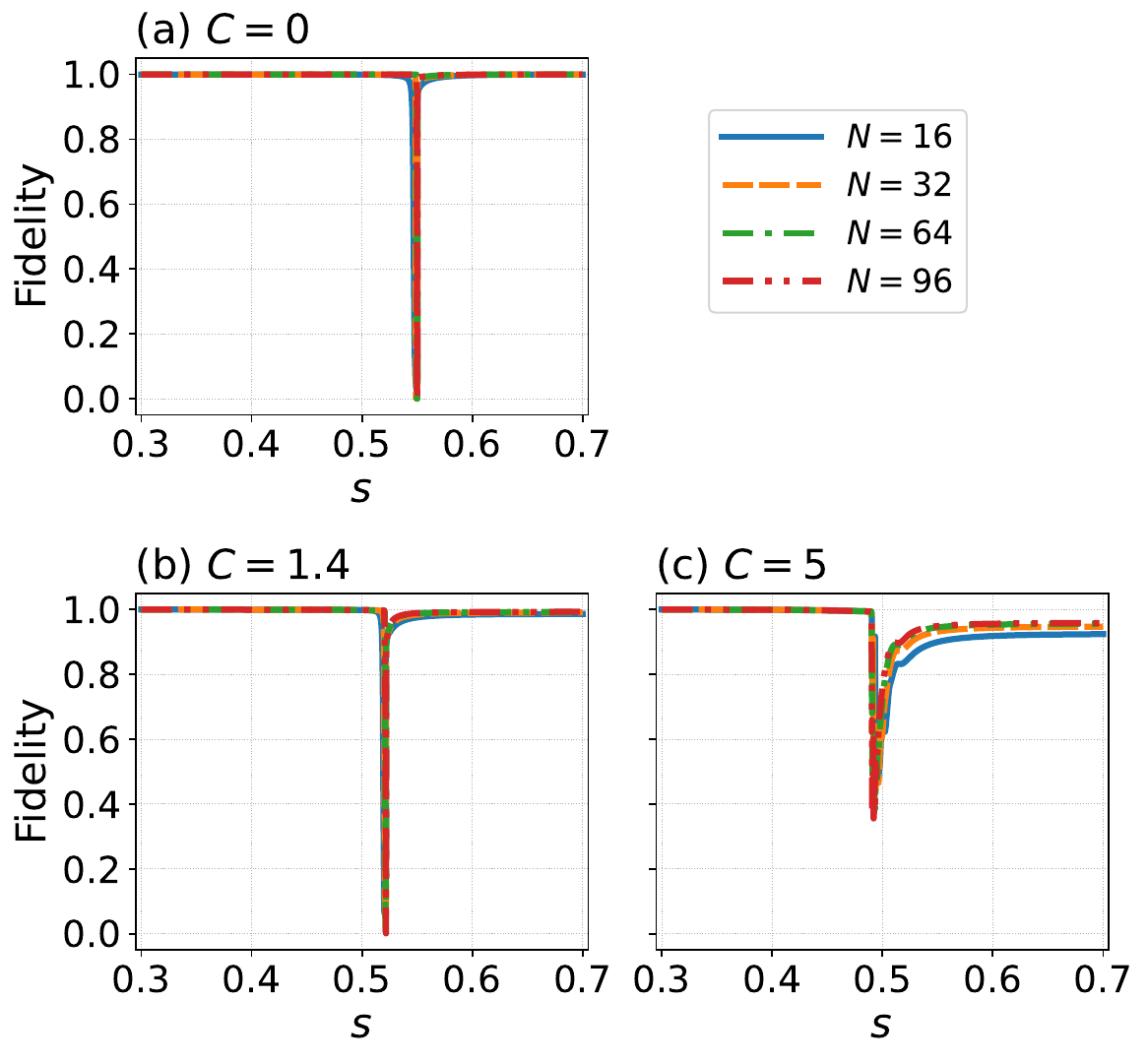}
\caption{The fidelity as a function of $s$ for $p=5$. $A(s)$ and $B(s)$ are from Eq. (\ref{eq:3}), with $A_0 = 3$, $\sigma^2=0.1$, and $B_0 = 40$.}
\label{fig:9}
\end{figure}

Let us consider the effect of the proposed nonstoquastic catalyst in a finite-size system to discuss the validity of the semiclassical approximation.
We can obtain the exact ground state $\ket{\psi_{\mathrm{GS}}}$ by diagonalizing the Hamiltonian (\ref{eq:8}). The details of the calculation are shown in Appendix \ref{app:C}.
We plot the order parameter for the system with $N=96$ spins, which is denoted as $m_{N=96}$, in Fig. \ref{fig:8}.
This result agrees with the semiclassical analysis shown in Figs. \ref{fig:2} and \ref{fig:3}.
Next, we calculate the fidelity between the semiclassical state $\ket{\psi_{\mathrm{SC, GS}}}$ and the exact ground state $\ket{\psi_{\mathrm{GS}}}$, namely, $|\bra{\psi_{\mathrm{SC, GS}}}\ket{\psi_{\mathrm{GS}}}|^2$.
The fidelity for $p=5$ is shown in Fig. \ref{fig:9}. 
We consider three values of $C$ and find the fidelity is almost $1$ except around the point of the phase transition in each case.
A possible reason for the sudden decrease in fidelity is that the location in $s$ for the phase transitions to occur in the finite-size calculation is slightly different from that in the semiclassical analysis.
These results suggest that the semiclassical state can reasonably approximate the exact ground state.

\begin{figure}[tp]
 \centering
  \includegraphics[width =0.4\textwidth]{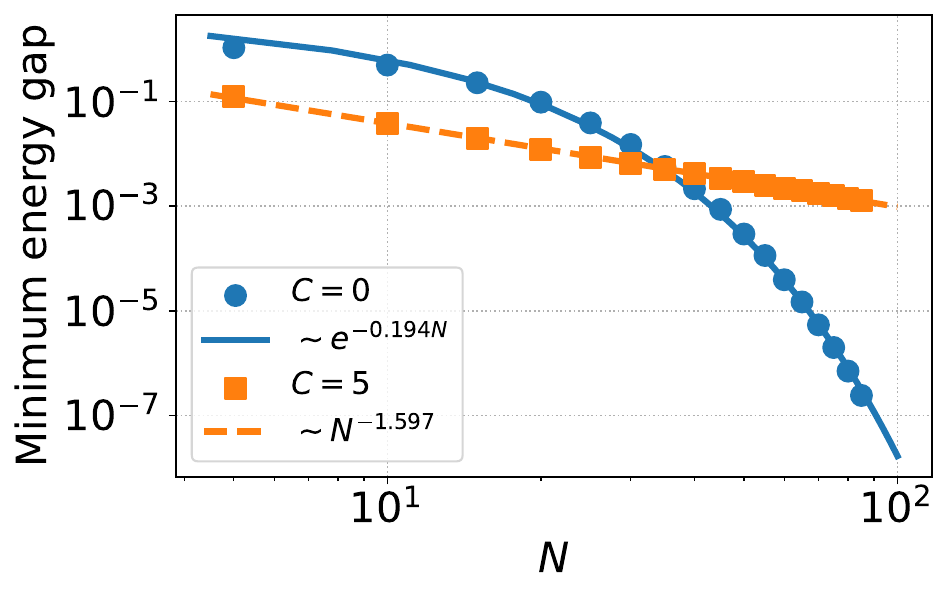}
\caption{The minimum energy gap against the system size $N$ for $p=5$. The blue circles and the orange squares are calculated from the eigenvalues in the stoquastic ($C=0$) and the nonstoquastic ($C=5$) cases, respectively. The blue solid curve and the orange dashed line show the results of exponential and polynomial fittings of the minimum energy gap for each case.}
\label{fig:10}
\end{figure}

By calculating the eigenvalues of the Hamiltonian (\ref{eq:8}), we can evaluate the minimum energy gap between the instantaneous ground state and the first excited state during annealing.
Figure \ref{fig:10} shows that the minimum energy gap exponentially (polynomially) decreases versus the system size $N$ in the stoquastic (nonstoquastic) case. 
For $N\geq 45$, the gap in the nonstoquastic case becomes larger than that in the stoquastic case.
Figure \ref{fig:10} is clear evidence that the proposed catalyst can qualitatively accelerate the present annealing protocol.

\section{Summary and Discussion}
\label{sec:6}
We considered the $p$-spin model in BQA described by the spin-1 operators and proposed a nonstoquastic catalyst.
We used the semiclassical analysis with the spin coherent state to investigate the phase transitions in stoquastic and nonstoquastic cases.
We found that, for specific cases, the proposed nonstoquastic catalyst (\ref{eq:6}) is effective for reducing the first-order phase transition appearing in the stoquastic case to the second-order one. This means that the catalyst will lead to performance improvement.
One needs a sufficient amplitude of the nonstoquastic catalyst and appropriate control of the driver Hamiltonian to remove the first-order phase transition.
Nevertheless, if the amplitude $C$ is small and the system has a first-order phase transition, the performance of BQA will be improved. As we saw in Fig. \hyperref[fig:3]{3(a)}, the jump in the order parameter at the first-order phase transition in the nonstoquastic case is less than the one in the stoquastic case.
The jump width corresponds to the energy-barrier width, which affects the probability that the system will reach the desired ground state thorough the quantum tunneling effect. Thus, a nonstoquastic catalyst would increase the probability and help with the ground-state search even when the first-order phase transition occurs.

We also evaluated the scaling of the energy gap by diagonalizing the Hamiltonian (\ref{eq:8}), and the results agree with our claim.
The fidelity shows that the semiclassical analysis is accurate enough to predict the exact ground state.
Since the semiclassical analysis is based on the static approximation, we could not evaluate the actual time to satisfy the adiabatic condition. 
However, as shown in a previous paper \cite{susa2017relation}, the semiclassical analysis provides a powerful tool to predict where the phase transition occurs, and we use it for the case of BQA.

We numerically calculated the endpoints of the curves for the first-order phase transitions in the phase diagram in Fig. \ref{fig:6}.
On the other hand, it was shown that for standard QA \cite{ohkuwa2017exact} the exact endpoints can be analytically derived for the mean-field model with the nonstoquastic catalyst.
Therefore, for our problem, we also might be able to predict the endpoints more precisely, which is left for future study. 

We note that the $p$-spin model with $p=4$ can be regarded as the mean-field approximation of the Lechner-Hauke-Zoller (LHZ) model with local four-body interactions \cite{lechner2015quantum,hartmann2019quantum}.
The LHZ model is considered practical architecture for BQA with KPOs to solve a fully connected problem Hamiltonian \cite{puri2017quantum,zhao2018two,kanao2021high}.
Thus, our results for the $p$-spin model will be helpful in designing real quantum devices.
However, to implement BQA with a nonstoquastic catalyst by using KPOs, we need to develop a framework in the bosonic system similar to our proposal. We also leave this for future work.

For further understanding of a nonstoquastic catalyst in BQA, it is desirable to study other instances, such as the weak-strong cluster problem, which also has a first-order phase transition \cite{albash2019role, takada2020mean}.
Additionally, developing other avenues for improving BQA is interesting.
Various approaches for improving standard QA have been studied, for example, inhomogeneous driving \cite{hartmann2019quantum, susa2018exponential, susa2018quantum}, reverse annealing \cite{ohkuwa2018reverse, yamashiro2019dynamics, passarelli2020reverse}, and counterdiabatic driving \cite{hartmann2019rapid,passarelli2020counterdiabatic, prielinger2021two,hartmann2022polynomial, passarelli2022optimal}.
Whether these are also applicable to BQA is an interesting topic.

\begin{acknowledgments}
The authors thank H. Nishimori for useful discussions and comments.
This paper is based on the results obtained from a project (Project No. JPNP16007) commissioned by the New Energy and Industrial Technology Development Organization (NEDO), Japan. 
This work is also supported by MEXT's Leading Initiative for Excellent Young Researchers and JST PRESTO (Grant No. JPMJPR1919), Japan.

\end{acknowledgments}

\appendix

\section{Quantum annealing with nitrogen vacancies centers in diamond}
\label{app:A}
Here, we discuss the experimental realization of BQA using the NV centers in diamonds.
The electronic ground state of the NV center is a spin triplet where we have $\ket{0}$ and $\ket{\pm 1}$.
We can polarize the NV center by applying a green laser \cite{gruber1997scanning}.
The state of the NV centers can be read out by measuring the photoluminescence~\cite{gruber1997scanning,jelezko2004observation} and can be controlled using microwave pulses~\cite{jelezko2004observation,de2010universal}.

The Hamiltonian of the NV centers is given as follows \cite{schirhagl2014nitrogen}:
\begin{align}
 H&=\sum _{i=1}^N D_i (\hat{S}_i^z)^2+ E_i\qty[(\hat{S}_i^x)^2 -(\hat{S}_i^y)^2] \notag \\
 &+2\lambda _i^{x} \hat{S}_i^x  \cos \omega _it +2\lambda _i^{y} \hat{S}_i^y  \cos \omega _it \notag\\
 &+\sum _{i,j}g_{ij}^z \hat{S}_i^z\hat{S}_j^z +g^{xy}_{ij} (\hat{S}_i^x\hat{S}_j^x +\hat{S}_i^y\hat{S}_j^y),
\end{align}
where $D_i$ denotes the zero-field splitting, $E_i$ denotes the strain, $\lambda _i^{x}$ ($\lambda _i^{y}$) denotes the Rabi frequency along the $x$ ($y$) direction, and $g^z$ ($g^{xy}$) denotes the coupling strength of the Ising (flip-flop) interaction.
The typical coupling strength between NV centers is around tens of kilohertz when the distance between NV centers is tens of nanometers \cite{dolde2013room,haruyama2019triple}.
We can control the zero-field splitting and strain by applying electric fields \cite{dolde2011electric}. We can determine the Rabi frequency by changing the microwave amplitudes
\cite{jelezko2004observation}. So we can set the energy scale of the NV centers to be tens of kilohertz when we implement the BQA.
By going to the rotating frame and using the rotating wave approximation, we obtain 
\begin{align}
 H\simeq& \sum _{i=1}^N (D_i-\omega _i) (\hat{S}_i^z)^2+ E_i\qty[(\hat{S}_i^x)^2
 -(\hat{S}_i^y)^2] \notag \\
 &+\lambda _i^{x} \hat{S}_i^x+\lambda _i^{y} \hat{S}_i^y+\sum _{i,j}g_z \hat{S}_i^z\hat{S}_j^z.
\end{align}
We can tune the value of $E_i$ by applying electric fields \cite{dolde2011electric}.
It is worth mentioning that, by adjusting the parameters with the NV centers, we can realize the Hamiltonian in Eq. \eqref{eq:5} with $p=2$.
It is possible to implement conventional QA with NV centers by using the spin lock technique
\cite{matsuzaki2020quantum,imoto2022obtaining}.
However, to perform the spin-lock, we need to perform single-qubit rotations. This means that the gate error will accumulate, and the success probability of QA will decrease.
On the other hand, when we perform BQA with the NV centers, we do not need to perform any gate operations. This shows the practical advantage of BQA with NV centers.

Let us discuss the possible realization of the nonstoquastic Hamiltonian (\ref{eq:8}) in our method.
It is known that the NV center and a bosonic mode can be coupled with either inductive or capacitive coupling \cite{bennett2013phonon,kubo2010strong,zhu2014observation,toida2016electron}.
Especially, we can couple a magnetic-field mode with the NV center in a subspace spanned by $|B\rangle $ and $|D\rangle $ \cite{doherty2012theory,matsuzaki2015improving,saijo2018ac}.
Within this subspace, the interaction with such a bosonic mode is $H_{\rm{I}}=\sum _i g(a \hat{\sigma}_i^+ + a ^{\dagger}\hat{\sigma}^{-}_i)$, where $\hat{\sigma}^+ =|B\rangle \langle D|$ and $\hat{\sigma}^- =|D\rangle \langle B|$ are the Pauli operators and $|B\rangle =(|+1\rangle +|-1\rangle )/\sqrt{2}$ [$|D\rangle =(|+1\rangle -|-1\rangle )/\sqrt{2}$] denotes a bright (dark) state.
In the dispersive regime where the detuning between the bosonic mode and the resonance frequency of the NV centers is much larger than $g$, we obtain 
$H_{\rm{I}} \propto (\sum _i \hat{\sigma}_i^z)^2$, where $\hat{\sigma}_i^z= |B\rangle \langle B|-|D\rangle \langle D|=|+1\rangle \langle -1|+|-1\rangle \langle +1|$~\cite{bennett2013phonon,tanaka2015proposed,dooley2016hybrid}. This corresponds to the nonstoquastic catalyst (\ref{eq:8}).

\section{Driver Hamiltonian rotated around $z$ axis}
\label{app:B}

\begin{figure*}[tbp]
\centering
  \includegraphics[width= 0.89\textwidth]{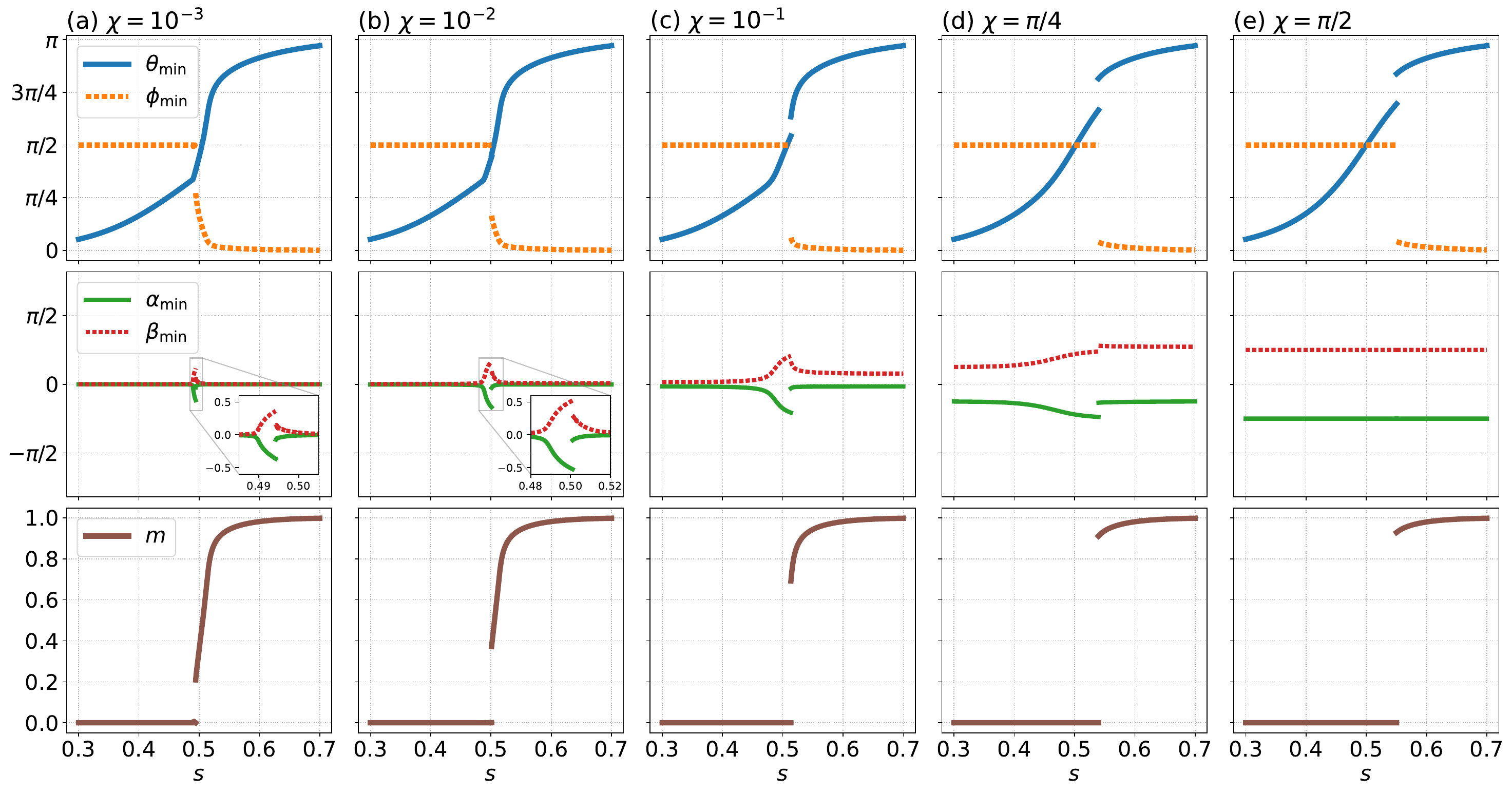}
  \includegraphics[width= 0.89\textwidth]{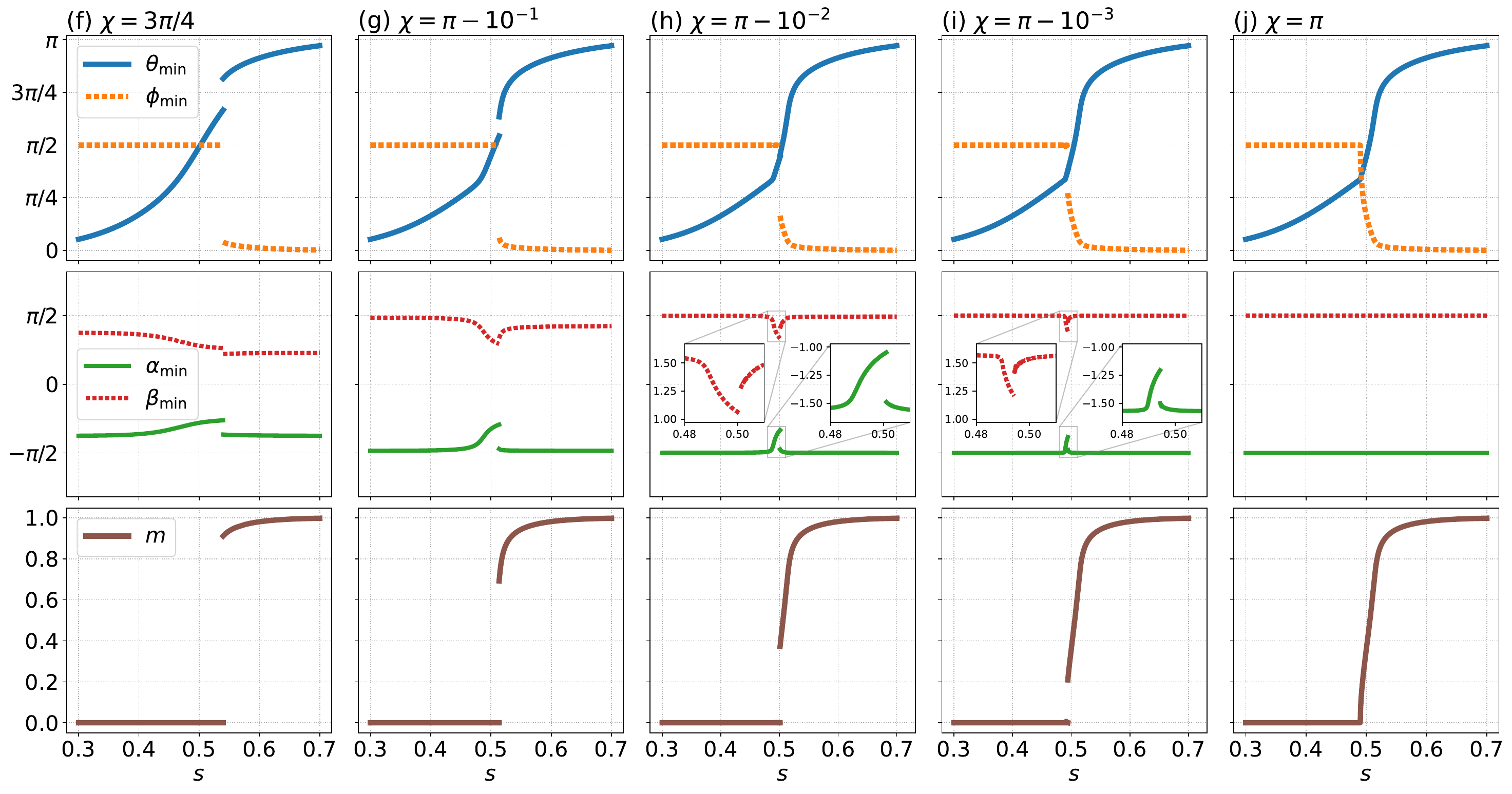}
\caption{Plots of $\theta_{\min}$, $\phi_{\min}$, $\alpha_{\min}$, $\beta_{\min}$, and order parameter $m$ against $s$ around the phase transitions for $p=5$ and $C=5$ for several $\chi$. In each case, $A(s)$ and $B(s)$ are from Eq. (\ref{eq:3}), with $A_0 = 3$, $\sigma^2=0.1$, and $B_0 = 40$.} 
\label{fig:11}
\end{figure*}

We consider the following Hamiltonian
\begin{align}
\label{eq:A1}
\hat{H}(s) =&\sum_{i=1}^N \qty[-A(s) \qty(\cos \frac{\chi}{2}\hat{S}_i^x+\sin\frac{\chi}{2}\hat{S}_i^y) -B(s)(\hat{S}_i^z)^2] \notag \\
&+CN\qty(\frac{1}{N}\sum_{i=1}^N (\hat{S}_i^x)^2-(\hat{S}_i^y)^2 )^2-N\qty(\frac{1}{N}\sum_{i=1}^N\hat{S}_i^z)^p,
\end{align}
where $\chi$ is the rotation angle of the driver Hamiltonian around the $z$ axis.
We note that the operator $\hat{S}_i^y$ has an imaginary component.
As in Sec. \ref{sec:4}, we calculate $\theta_{\min},\phi_{\min},\alpha_{\min},\beta_{\min}$, and the order parameter $m$ from the semiclassical potential of the Hamiltonian (\ref{eq:A1}). 
The semiclassical potential is as follows:
\begin{widetext}
\begin{align}
\label{eq:A2}
V_{\textrm{SC}}(s,\chi.\theta,\phi,\alpha,\beta) 
=& \lim_{N \rightarrow \infty} \frac{1}{N}\bra{\psi_{\textrm{SC}}}\hat{H}(s)\ket{\psi_{\textrm{SC}}} =-\frac{A(s)}{\sqrt{2}}\sin \theta \qty[\cos\frac{\phi}{2} \cos\qty(\alpha+\frac{\chi}{2}) +\sin\frac{\phi}{2}\cos\qty(\beta-\frac{\chi}{2})]\notag \\
&-B(s)\sin^2\frac{\theta}{2}+C\qty(\sin^2\frac{\theta}{2}\sin \phi \cos(\alpha-\beta) )^2 -\qty(\sin^2\frac{\theta}{2} \cos \theta )^p.
\end{align}
\end{widetext}
Here, we fix $p=5$. In the stoquastic case $C=0$, $\theta_{\min}$, $\phi_{\min}$ and the order parameter $m$ are the same as in Fig. \ref{fig:2}, while $\alpha_{\min} = -\chi/2$ and $\beta_{\min} = \chi/2$. 
Next, we consider the nonstoquastic case $C=5$.
Figure \ref{fig:11} shows the numerical results for several $\chi$.
Note that $\chi=0$ reproduces the result in Fig. \hyperref[fig:3]{3(b)}, where the first-order phase transition disappears.
However, as indicated in Fig. \hyperref[fig:11]{11(a)}, the first-order phase transition reappears even if $\chi$ rotates slightly from zero.
The top half of Fig. \ref{fig:11} shows that the jump in the order parameter increases as $\chi$ increases from $0$ to $\pi/2$.
We also find $\alpha_{\min}$ and $\beta_{\min}$ deviate from zero by rotating $\chi$.
In Fig. \hyperref[fig:11]{11(e)}, we obtain $\alpha_{\min}=-\pi/4$ and $\beta_{\min}=\pi/4$.
This means that the third term, given by the nonstoquastic catalyst, in the semiclassical potential (\ref{eq:A2}) vanishes. Therefore, the nonstoquastic catalyst is ineffective for the case of $\chi=\pi/2$.

The bottom half of Fig. \ref{fig:11} shows that the jump in the order parameter decreases as $\chi$ increases from $\pi/2$ to $\pi$.
Figure \hyperref[fig:11]{11(j)} for $\chi = \pi$ shows the second-order phase transition with $\alpha_{\min}=-\pi/2$ and $\beta_{\min}=\pi/2$, where the third term in the semiclassical potential (\ref{eq:A2}) remains.
However, the first-order phase transition shows up even if $\chi$ shifts slightly from $\pi$, as shown in Figs. \hyperref[fig:11]{11(g)}-\hyperref[fig:11]{11(i)}.
Therefore, we need to take care with the rotation angle of the driver term around the $z$ axis when we remove the first-order phase transition using the proposed nonstoquastic catalyst.

\section{Diagonalizing the Hamiltonian (\ref{eq:8})}
\label{app:C}
To evaluate the exact ground state for a finite-size system, we consider the matrix representation of the total Hamiltonian (\ref{eq:8}).
Since the Hamiltonian is symmetric under the spin permutation, we can restrict our computation to the symmetric subspace. Hence, the basis can be described by the number state for $N$ spins, defined as
\begin{align}
\ket{n_1, n_0, n_{-1}} =& \sqrt{\frac{n_1!n_0!n_{-1}!}{N!}} \sum \mathcal{P}\qty {\ket{1}^{\otimes n_1} \ket{0}^{\otimes n_0} \ket{-1}^{\otimes n_{-1}}},
\end{align}
where $n_1$, $n_0$, and $n_{-1}$ denote the number of spins in states $\ket{1}$, $\ket{0}$, and $\ket{-1}$, respectively, and $\sum \mathcal{P}$ represents the sum over all permutations of $N$ entries.
We then obtain the matrix element of the Hamiltonian
$\bra{n_1,n_0,n_{-1}}\hat{H}(s)\ket{n_1,n_0,n_{-1}}  = [\hat{H}(s)]_{(n_1,n_0,n_{-1}),(n_1^{\prime},n_0^{\prime},n_{-1}^{\prime})}$ for all possible combinations of $(n_1,n_0,n_{-1})$ as 
\begin{align}
&[\hat{H}(s)]_{(n_1,n_0,n_{-1}),(n_1,n_0,n_{-1})} \notag \\
=&  -B(s) (N-n_0)+\frac{C}{N}( 2n_1n_{-1} + n_1+n_{-1}) \notag \\
&-N\qty(\frac{n_1-n_{-1}}{N})^p, \\
&[\hat{H}(s)]_{(n_1,n_0+1,n_{-1}-1),(n_1,n_0,n_{-1})} \notag \\ 
=& [\hat{H}(s)]_{(n_1,n_0,n_{-1}),(n_1,n_0+1,n_{-1}-1)} \notag \\ 
=& [\hat{H}(s)]_{(n_1+1,n_0-1,n_{-1}),(n_1,n_0,n_{-1})} \notag \\ 
= &[\hat{H}(s)]_{(n_1,n_0,n_{-1}),(n_1+1,n_0-1,n_{-1})}  \notag \\ 
=& -A(s)\sqrt{\frac{(n_0+1)n_{-1}}{2}},  \\
&[\hat{H}(s)]_{(n_1+2,n_0,n_{-1}-2),(n_1,n_0,n_{-1})} \notag \\ 
=&[\hat{H}(s)]_{(n_1,n_0,n_{-1}),(n_1+2,n_0,n_{-1}-2)} \notag \\ 
=&\frac{C}{N} \sqrt{(n_1+2)(n_1+1)n_{-1}(n_{-1}-1)}.
\end{align}

By diagonalizing the matrix, we obtain the eigenvalues $e_{n_1,n_0,n_{-1}}^{\mathrm{GS}}$ of the exact ground state of the finite system as
\begin{align}
\ket{\psi_{\mathrm{GS}}^{(N)} } = \sum_{n_1 + n_0+n_{-1} = N}  e_{n_1,n_0,n_{-1}}^{\mathrm{GS}} \ket{n_1,n_0,n_{-1}}.
\end{align}
From the eigenvalues, we can calculate the energy gap and also the order parameter for the finite-size system as
\begin{align}
m_{N} :=& \frac{1}{N} \bra{\psi_{\mathrm{GS}}^{(N)}} \sum_{i=1}^N \hat{S}_i^z \ket{\psi_{\mathrm{GS}}^{(N)}} \notag \\
=& \frac{1}{N}\sum_{n_1 + n_0+n_{-1} = N}  (n_1-n_{-1})|e_{n_1,n_0,n_{-1}}|^2.
\end{align}
Finally, we can calculate the inner product between the spin-coherent ground state and the exact ground state as 
\begin{align}
\bra{\psi_{\mathrm{SC, GS}}}\ket{\psi_{\mathrm{GS}}^{(N)}}
=& \sum_{n_1 + n_0+n_{-1} = N}  e_{n_1,n_0,n_{-1}}^{\mathrm{GS}}\sqrt{\frac{N!}{n_1!n_0!n_{-1}!}} \notag \\
&\times \qty(\sin \frac{\theta_{\min}}{2} \cos\frac{\phi_{\min}}{2})^{n_1} \qty(\cos\frac{\theta_{\min}}{2})^{n_0} \notag \\
&\times \qty(\sin \frac{\theta_{\min}}{2} \sin\frac{\phi_{\min}}{2})^{n_{-1}}.
\end{align}
In this derivation, we have fixed $\alpha_{\min}=\beta_{\min}=0$.
We can then calculate the fidelity.

\end{document}